\documentclass[12pt,a5paper]{article}

\usepackage{amsmath}
\allowdisplaybreaks
\IfFileExists{ajr.sty}{\usepackage{ajr}}{}
\usepackage{defs}
 
\begin{document}

\title{A normal form of thin fluid film equations provides initial
conditions}

\author{A.~J.~Roberts\thanks{Department of Mathematics and
Computing, University of Southern Queensland, Toowoomba~4352,
\textsc{Australia}.  \protect\url{mailto:aroberts@usq.edu.au}}}

\maketitle

\begin{abstract}
	We use dynamical systems theory to construct the normal form of the
	Navier--Stokes equations for the flow of a thin layer of fluid upon
	a solid substrate.  The normal form equations illuminate the fluid
	dynamics by decoupling the long-term flow of interest from the
	rapid viscous decay of the transient shear modes.  The normal form
	clearly shows the centre manifold of the lubrication model and
	shows the result that the initial condition for the fluid thickness
	of the lubrication model is \emph{not} the initial physical fluid
	thickness, but instead is modified by the initial lateral shear
	flow.  With these initial conditions, better forecasts will be made
	using the lubrication model.  This dynamical systems approach
    will also enable similar illumination of other complicated
    models of dynamics.
\end{abstract}

\tableofcontents
   
\section{Introduction}
\label{intro}
The flow of a thin layer of fluid is important in numerous industrial
and natural processes: industrial applications include the coating
processes of autobodies, beverage containers, sheet goods and films,
decorative coating, and gravure roll coating; biomedical applications
include the liquid films covering the cornea of the eye or protecting
the linings of the lungs~\cite{Grotberg94}.  A variety of mathematical
models of the fluid dynamics have been developed to aid understanding
and prediction of the flow.  For examples see the extensive reviews of
Chang~\cite{Chang94}, Ruschak~\cite{Ruschak85} or Oron, Davis and
Bankoff~\cite{Oron97}.  On a flat substrate and in one lateral
dimension the usual model for the evolution of a viscous fluid film's
thickness~$\eta(x,t)$, driven only by surface tension, is the
following nondimensional ``lubrication'' model~\cite[e.g.]{Roy96}
\begin{equation}
    \D \eta t\approx -\rat13\we\D{}x\left(\eta^3\Dn\eta{x}3\right)\,,
    \label{eq:cmmod}
\end{equation}
where the Weber number~$\we$ measures the strength of surface tension
in the thin film flow.  This is a considerable simplification of the
governing equations when compared to the full Navier--Stokes equations.

Consider making a forecast with the lubrication model~(\ref{eq:cmmod})
from some initial state of the fluid, say with physical
thickness~$\eta_0$.  Surely one just sets the initial field~$\eta(x,0)$
to the initial thickness of the actual fluid~$\eta_0(x)$.  Herein we
discover this is incorrect and that instead the correct initial
condition for the lubrication model~(\ref{eq:cmmod}) is more accurately
\begin{equation}
    \eta(x,0)\approx \eta_0
    -\re\D{}x\int_0^{\eta_0} (\eta_0 y-\half y^2)u_0 \,dy\,,
    \label{eq:cmic}
\end{equation}
where $u_0(x,y)$ is the initial lateral velocity field of the fluid,
and $\re$ is an appropriate Reynolds number.  This initial condition
accounts for the flow of fluid in the transient viscous decay of the
initial velocity field.

For example, two colliding sheets of fluid form a hump.  Suppose the
initial fluid state is that of a constant thickness, $\eta_0$~constant,
and Couette flow, $u_0=U_0f(x)y/\eta_0$ where $f(x)$ describes the
lateral variations in strength of the initial lateral shear flow.
Then from~(\ref{eq:cmic}) the corresponding initial thickness for the
model~(\ref{eq:cmmod}) is
\begin{equation}
    \eta(x,0)\approx \eta_0 -\re\frac{5 U_0\eta_0^3}{24}f'(x)\,.
    \label{eq:cmicf}
\end{equation}
Two colliding sheets of fluids corresponds to a localised and negative
derivative~$f'(x)$; in this case the initial condition~(\ref{eq:cmicf})
says the best forecast is obtained by starting the model from an
initial condition with a hump proportional to~$-f'(x)$.  This hump
accounts for the transient collection of fluid at the collision as the
lateral velocity decays through viscosity.  This physically reasonable
result agrees with earlier work~\cite{Suslov99, Suslov98b}.

Here we develop analysis, Section~\ref{SScon}, to determine that the
parabolic weight function is correct in the integral
of~(\ref{eq:cmic}).  Further, we introduce and use a framework,
Section~\ref{Sflow}, to show that~(\ref{eq:cmic}) is just the first
approximation in an asymptotic series for the correct prescription of
the initial conditions for the lubrication model~(\ref{eq:cmmod}), see
Section~\ref{Srev}.  The theoretical framework is that of the normal
form transformation in dynamical systems with centre
manifolds~\cite{Elphick87b, Cox93b}.  The continuity equation of
incompressible fluid dynamics is an algebraic equation; thus, we make
the novel application of the normal form transformation to differential
algebraic systems.

Because of the complicated detail and form of the fluid dynamics
equations, we introduce the normal form transformation in
Section~\ref{smpex} using a relatively simple example dynamical system.
The transformation is a near identity transformation with the aim of
nonlinearly decoupling the dynamical equations as far as possible.  In
systems with a centre manifold, the centre manifold model appears
immediately, see the example of~Section~\ref{sec:nfint}, as first recognised
by Elphick et al.~\cite{Elphick87b}.  Cox and Roberts~\cite{Cox93b}
then recognised that the normal form also immediately provides the
correct initial conditions for a model, see the example
of~Section~\ref{SSproj}.  The example is a coupled differential algebraic
system; in Sections~\ref{sec:nfcon}--\ref{Shomo} we show that the
normal form may none-the-less be constructed in the usual manner.

Then in Section~\ref{Sflow} we introduce the fluid equations and
develop the normal form transformation that decouples the viscously
decaying dynamics of lateral shear from the long lasting slow evolution
of the thickness of the fluid film.  When decoupled, the arguments of
Section~\ref{Srev} provide initial conditions for the lubrication
model~(\ref{eq:cmmod}).  Equation~(\ref{eq:cmic}) gives the leading
nontrivial terms in the asymptotic expression for the initial
condition.

\section{An example introduces the normal form}
\label{smpex}

Here we explore a straightforward artificial example (introduced
in~\cite{Suslov98b}) to illustrate the techniques used later in the
normal form derivation of the initial conditions~(\ref{eq:cmic}) for
the lubrication model~(\ref{eq:cmmod}).  This example reflects in
detail many essential aspects of the relation between the normal form
and initial conditions, and how to use the freedom inherent in the
derivation.  Someone very familiar with normal form transformations
may skip this section.

\begin{figure}
    \centering
    \begin{tabular}{cc}
        \raisebox{23ex}{$u$} &  
        \includegraphics[width=0.9\textwidth]{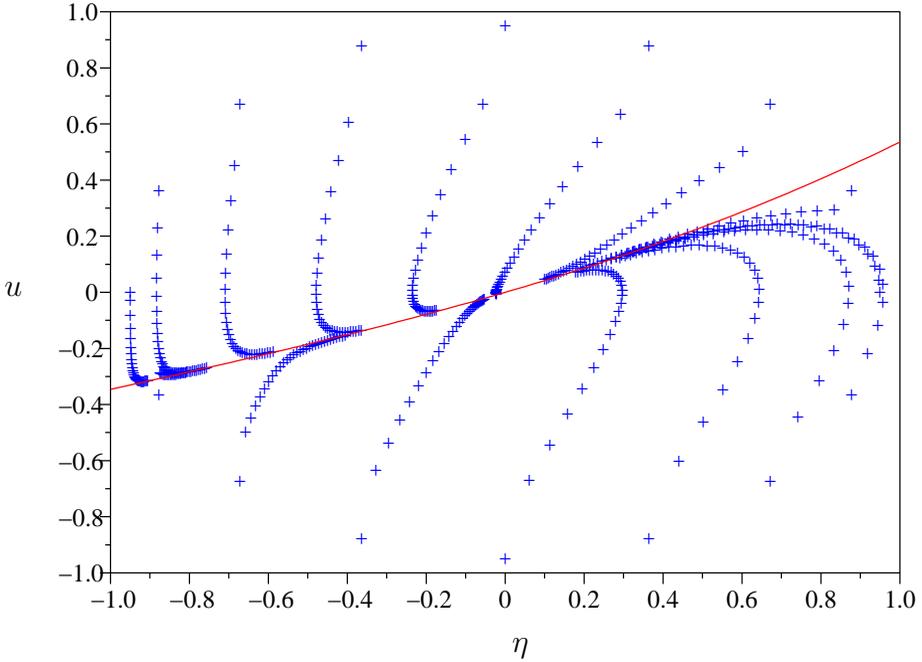} \\
         & $\eta$  \\
    \end{tabular}
	\caption{simulations of the example system~(\ref{seu}--\ref{sek})
	for $\epsilon=0.4$ over time interval~$6$ at time steps of $\Delta
	t=0.2$ and projected onto the $\eta u$-plane.  This shows the
	exponential quick collapse of states onto the low-dimensional
	centre manifold (solid curve) given by~(\ref{su}).}
    \label{fig:se}
\end{figure}

Consider the following index~$2$ differential-algebraic system which is
structurally analogous to the thin fluid film
equations~(\ref{nseq}--\ref{svekc}) introduced in Section~\ref{Sflow}
(variable subscripts denote differentiation):
\begin{eqnarray}
  &&u_t+\epsilon u^2+uv+\epsilon p+\left(1+\epsilon^2\right)u
  +\epsilon\eta=0\,,\label{seu}\\
  &&v_t+\epsilon uv+v^2+p+\left(1+\epsilon^2\right)v=0\,,
  \label{sev}\\
  &&\quad\quad\epsilon u+v=0\,,
  \label{sec}\\
  &&\eta_t+\epsilon u\eta-v=0\,.\label{sek}
\end{eqnarray}
Some trajectories of solutions are shown in Figure~\ref{fig:se}.  In
this section $u(t)$, $v(t)$, $p(t)$ and~$\eta(t)$ represent analogues
of a fluid film's velocity, pressure and thickness.  The small
parameter~$\epsilon$ is introduced to mimic slow lateral variations in
the fluid thickness and flow.  The normal form transformation we
explore in this section clarifies the long term evolution of the above
finite set of ordinary differential evolution equations.

\paragraph{Low-dimensional dynamics:}
When~$\epsilon=0$\,, the system (\ref{seu}--\ref{sek}) has the
equilibrium solutions $\eta=\mbox{const}$, $u=v=p=0$ (mimicking a
uniform motionless fluid layer of thickness~$\eta$).  These equilibria,
call the set of them~$\cM_0$, are stable: linearising, putting
$\epsilon=0$ and seeking nontrivial solutions proportional
to~$\exp(\lambda t)$ leads to requiring the determinant
\begin{equation}
    \left| \begin{array}{cccc}
        \lambda+1&0&0&0\\
        0&\lambda+1&0&0\\
        0&1&1&0\\
        0&-1&0&\lambda
    \end{array} \right|
    =\lambda(\lambda+1)=0\,.
    \label{eq:det}
\end{equation}
This shows: the rapid $\Ord{e^{-t}}$~decay to any of the equilibria is
determined by the $u$-equation~(\ref{seu}); equation~(\ref{sek})
determines $\eta$~is a neutral mode; whereas the other two equations do
not contribute dynamics because of the need to satisfy the algebraic
`continuity' equation~(\ref{sec}).  For small $\epsilon\ne0$\,, instead
of $\eta$~being constant, we thus expect that~$\eta$ will evolve slowly
on a nearby and similarly exponentially attractive set of states,
called the centre manifold.  This is confirmed by the numerical
simulations shown in Figure~\ref{fig:se}.  We soon,
Section~\ref{sec:nfint}, confirm this nearby centre manifold of slow
long term evolution is
\begin{eqnarray}
  u&=&-\epsilon \eta \left(1-3\epsilon^2\right)+\epsilon^3\eta ^2
  +\Ord{\epsilon^5}\,,\label{su}\\
  v&=&\phantom{-}\epsilon^2\eta \left(1-3\epsilon^2\right)-\epsilon^4\eta ^2
  +\Ord{\epsilon^5}\,,\label{sv}\\
  p&=&-\epsilon^2\eta \left(1-\epsilon^2\right)
  +\Ord{\epsilon^5}\,.\label{sp}
\end{eqnarray}
On this centre manifold the slow model evolution is
\begin{equation}
    \eta_t=\epsilon^2(1+\eta)\eta
    -\epsilon^4 (1+\eta)(3+\eta) 
     +\Ord{\epsilon^5}\label{se}\,.
\end{equation}
By the centre manifold relevance theorem~\cite[Chapt.~5,
e.g.]{Kuznetsov95} or~\cite{Carr81}, we expect the long term behaviour
of all nearby solutions are described by~(\ref{su}--\ref{se}).

Elphick et~al.~\cite{Elphick87b} noted that low-dimensional centre
manifold models such as~(\ref{su}--\ref{se}) are an immediate
consequence of transforming differential equations into a normal form.
Cox \& Roberts~\cite{Cox93b} observed that the projection of initial
conditions onto such models also immediately follows from such a normal
form.  However, neither group explicitly addressed dynamics governed by
systems of differential-algebraic equations.  Here we demonstrate how
the techniques adapt easily to differential-algebraic systems such as
both~(\ref{seu}--\ref{sek}) and also incompressible fluid dynamics, see
Section~\ref{Sflow}.  We illuminate the dynamics and modelling of such
systems with a normal form transformation.

\subsection{Interpret the normal form transformation}
\label{sec:nfint}
In Section~\ref{sec:nfcon} we will show how to construct the \emph{near
identity} transform to new dynamical variables that simplifies the
description of the dynamics.  But first we show how the resultant
transformation illuminates the dynamics by decoupling the long-term
evolution from the short-term decaying transients, and by decoupling
the algebraic component of the governing equations.  Substitute into
the dynamical system~(\ref{su}--\ref{se}), with new variables denoted
by Fraktur font, the near identity transformation:
\begin{eqnarray}
    \eta &=& \hh +\epsilon(1+\hh)\uu +\epsilon^2\half (1+\hh)\uu^2 
    +\epsilon^3(1+\hh)[-(2+\hh)\uu +\rat16\uu^3]
    \nonumber\\&&\phantom{\hh}
    +\epsilon^4 (1+\hh)[-(\rat74+\rat34\hh)\uu^2 +\rat1{24}\uu^4  ]
    +\Ord{\epsilon^5}\,,
    \label{eq:tnfh}
    \\
    u &=& \uu- \epsilon \hh +\epsilon^3[ (3+\hh)\hh +\half(1+\hh)\uu^2 ]
    +\epsilon^4\rat1{12}(1+\hh)\uu^3
    +\Ord{\epsilon^5}\,,
    \label{eq:tnfu}
    \\
    v &=& \vv-\epsilon\uu+\epsilon^2\hh 
    -\epsilon^4[ (3+\hh)\hh +\half(1+\hh)\uu^2 ]
    +\Ord{\epsilon^5}\,,
    \label{eq:tnfv}
    \\
    p &=& \pp-\epsilon^2\hh -\epsilon^3(1+\hh)\uu
    +\epsilon^4[ \hh-\half(1+\hh)\uu^2 ]
    +\Ord{\epsilon^5}\,.
    \label{eq:tnfp}
\end{eqnarray}
Then the example dynamical system~(\ref{seu}--\ref{sek}) becomes
\begin{eqnarray}&&
    \hh_t=\epsilon^2(1+\hh)\hh
    -\epsilon^4 (1+\hh)(3+\hh) 
     +\Ord{\epsilon^5}\,,
     \label{eq:nfh}
     \\&&
    \uu_t=-\left[\ 1+\epsilon^2(2+\hh) -\epsilon^4(1+\hh)(3+\hh)
    \ \right]\uu 
    +\Ord{\epsilon^5}\,,
    \label{eq:nfu}
    \\&&
    \vv=\pp=0\,.
    \label{eq:nfy}
\end{eqnarray}
Crucially, \emph{the normal form system~(\ref{eq:nfh}--\ref{eq:nfy})
captures all the solutions of the original
system~(\ref{seu}--\ref{sek})}, at least in some neighbourhood of the
equilibria~$\cM_0$.  The reason is simply that the
transformation~(\ref{eq:tnfh}--\ref{eq:tnfp}) is a smooth
reparametrisation of the complete state space near~$\cM_0$.  Thus from
arbitrary feasible initial conditions the normal form
system~(\ref{eq:nfh}--\ref{eq:nfy}) retains all the transient dynamics
and all the long-term dynamics provided the dynamics stay within the
neighbourhood of~$\cM_0$.  Consider the effects of this
transformation.
\begin{enumerate}
	\item The algebraic part of the original
	system~(\ref{seu}--\ref{sek}) is transformed into the trivial and
	decoupled algebraic equations~(\ref{eq:nfy}), namely $\vv=\pp=0$\,.
    
	\item See immediately from the form~(\ref{eq:nfu}) that $\uu=0$ is
	an invariant manifold of the dynamics, and is exponentially quickly
	attractive to a wide variety of nearby initial conditions.  Thus
	this normal form clearly displays the attractive centre manifold is
	$\uu=0$\,.  In original variables, the
	transformation~(\ref{eq:tnfh}--\ref{eq:tnfp}) then immediately
	shows the centre manifold maps to $\eta=\hh$ and the earlier
    claimed~(\ref{su}--\ref{sp}).
   
    \item Also see the evolution~(\ref{eq:nfh}) of~$\hh$
	reaffirms the earlier claimed centre manifold model~(\ref{se}).

	\item Lastly, a crucial feature of the
	$\hh$~evolution~(\ref{eq:nfh}) is that it is independent of~$\uu$
	in this normal form.  Thus all states with the same~$\hh$ but
	different~$\uu$ only differ in the evolution of~$\uu$, the
	$\hh$~evolution is identical.  Next we use this to deal with
    initial conditions.
\end{enumerate}

\subsection{Project initial conditions}
\label{SSproj}

Consider the result of specifying some initial condition in a
simulation, say
\begin{equation}
    \eta(0)=\eta_0\,,\quad
    u(0)=u_0\,,\quad
    v(0)=v_0\,,\quad
    p(0)=p_0\,.
    \label{eq:nfic}
\end{equation}
In general these will not lie on the centre manifold
model~(\ref{su}--\ref{sp}), see for example the simulations shown in
Figure~\ref{fig:se}.  Following the arguments in~\cite{Cox93b}, we use
the normal form to deduce how to project such an initial condition into
one for the low-dimensional model~(\ref{se}).

\begin{figure}
    \centering
    \begin{tabular}{cc}
        \raisebox{23ex}{$u$} &  
        \includegraphics[width=0.9\textwidth]{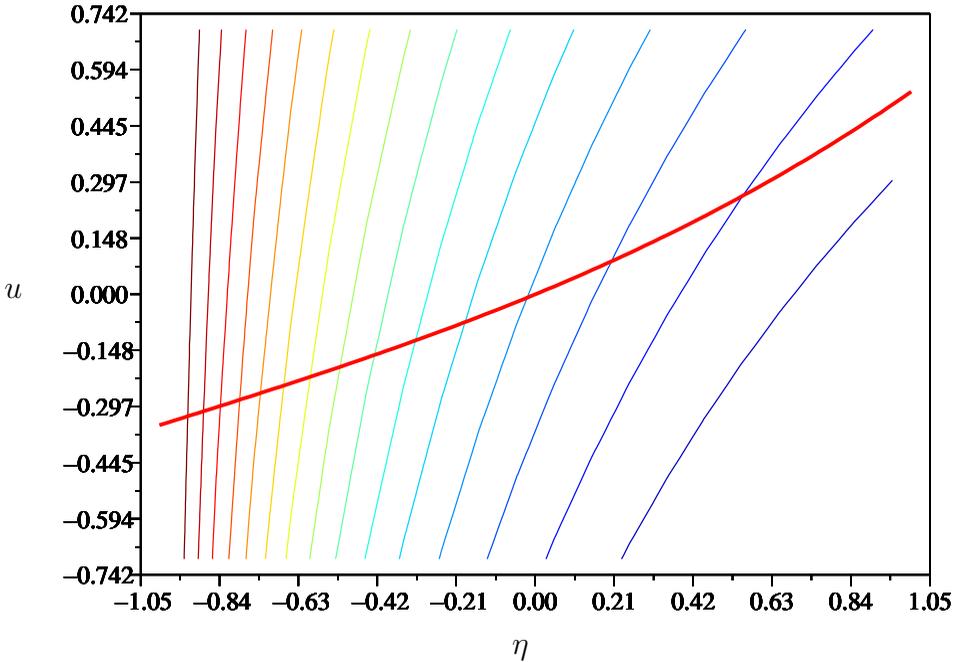} \\
         & $\eta$  \\
    \end{tabular}
	\caption{centre manifold $\uu=0$ (thick curve) of the example
	system~(\ref{seu}--\ref{sek}) for $\epsilon=0.4$\,.  The
	16~transverse curves, called isochrons, show initial points which
	have the same long term dynamics.}
    \label{fig:seiso}
\end{figure}
But first we numerically illustrate the projection.  Numerical
solutions are feasible for simple systems.  By running simulations,
such as those in Figure~\ref{fig:se}, we draw contours of all initial
conditions whose subsequent solutions end up at the same location (to
an exponentially small difference) on the centre manifold after some
fixed time, see Figure~\ref{fig:seiso}.\footnote{These contours are
called isochrons~\cite{Guckenheimer75, Cox93b}; they are also known as
the stable fibrations~\cite[\S5.2]{Murdock03}.} All the points reaching
the same location at the same fixed time must then have the same long
term evolution.  Hence, all initial conditions lying on one of the
contours in Figure~\ref{fig:seiso} must be projected onto the centre
manifold where the contour intersects the manifold.  Then the model
will predict the correct long term evolution from the originally
specified initial condition.

Any such initial condition corresponds to some
point~$(\hh_0,\uu_0,\vv_0,\pp_0)$ in the transformed variables.  First,
revert the series in the transformation~(\ref{eq:tnfh}--\ref{eq:tnfp})
to find the initial state in the normal form variables
\begin{eqnarray}
    \hh_0&=&\eta_0 -\epsilon(1+\eta_0)u_0
    -\epsilon^2(1+\eta_0)(\eta_0-\half u_0^2)
    \nonumber\\&&{}
    +\epsilon^3\rat16(1+\eta_0)(18+18\eta_0-u_0^2)u_0
    \nonumber\\&&{}
    +\epsilon^4\rat1{24}(1+\eta_0)(84\eta_0^2 +144\eta_0
    -90\eta_0u_0^2 -102u_0^2 +u_0^4)
    +\Ord{\epsilon^5}\,,\label{eq:nfih}\\
    \uu_0&=&u_0 +\epsilon \eta_0 -\epsilon^2(1+\eta_0)u_0
    -\epsilon^3(4+2\eta_0)\eta_0
    \nonumber\\&&{}
    +\rat14\epsilon^4(1+\eta_0)(24 +u_0^2 +16\eta_0 )u_0
    +\Ord{\epsilon^5}\,,\label{eq:nfiu}
    \\
    \vv_0&=&v_0+\epsilon u_0\,, \label{eq:nfiv}
    \\
    \pp_0&=&p_0+\epsilon^2\eta_0-\epsilon^4\eta_0
    +\Ord{\epsilon^5}\,. \label{eq:nfip}
\end{eqnarray}
Second, for this system with an algebraic component the initial
conditions~(\ref{eq:nfic}) must be consistent with the algebraic
constraints~(\ref{eq:nfy}), namely $\vv_0=\pp_0=0$\,; hence, we require
the original system to satisfy
\begin{equation}
    v_0=-\epsilon u_0
    \quad\mbox{and}\quad
    p_0=-\epsilon^2 \eta_0 +\epsilon^4 \eta_0+\Ord{\epsilon^5}\,.   
\end{equation}
Finally, when $\uu_0$ is non-zero the initial condition is off the
centre manifold $\uu=0$\,.  But in the normal
form~(\ref{eq:nfh}--\ref{eq:nfu}) the evolution of~$\hh$ is unaffected
by the evolution of~$\uu$.  Further, $\uu$ decays to zero exponentially
quickly.  Thus, apart from an exponentially decaying transient, the
evolution from~$(\hh_0,\uu_0)$ will be identical to that from the
point~$(\hh_0,0)$ on the centre manifold.  Thus the appropriate initial
condition for the low-dimensional model~(\ref{se}) is not
$\eta(0)=\eta_0$ but instead that $\eta(0)=\hh_0$ as given
by~(\ref{eq:nfih}).  This projection of the initial condition takes
into account the initial transients in the dynamics.

\subsection{Construct the normal form}
\label{sec:nfcon}
We now explore how to construct the particular normal form
transform~(\ref{eq:tnfh}--\ref{eq:tnfp}) that has the requisite
properties we used in the previous Section~\ref{SSproj}.  The
construction has considerable detail, see~\cite{Cox93b, Vallier93,
Murdock03}, with many subtleties in its application to this problem. 
To find the transformation~(\ref{eq:tnfh}--\ref{eq:tnfp}) with
corresponding evolution~(\ref{eq:nfh}--\ref{eq:nfy}), we seek a
\emph{near identity} coordinate transform in the general form
\begin{eqnarray}
    && 
    \eta=\hh+\fh(\hh,\uu,\epsilon)\,,\quad
    u=\uu+\fu(\hh,\uu,\epsilon)\,,
    \nonumber\\&&
    v=\vv+\fv(\hh,\uu,\epsilon)\,,\quad
    p=\pp+\fp(\hh,\uu,\epsilon)\,,
    \label{eq:nft}
\end{eqnarray}
such that the differential-algebraic system~(\ref{seu}--\ref{sek})
takes this differential-algebraic normal form
\begin{equation}
    \hh_t=\gh(\hh,\epsilon)\,,\quad
    \uu_t=-\uu+\gu(\hh,\uu,\epsilon)\,,\quad
    \vv=\pp=0\,,
    \label{eq:nfe}
\end{equation}
where the functions $\fh$, $\fu$, $\fv$, $\fp$, $\gh$~and~$\gu$ are
strictly nonlinear functions of $\hh$, $\uu$~and~$\epsilon$, and where
we require
\begin{equation}
    \gu\mbox{ strictly linear in }\uu\,,
    \quad\mbox{that is }\gu\propto\uu\,.
    \label{eq:zgu}
\end{equation}
That these properties~(\ref{eq:nfe}--\ref{eq:zgu}) can be found is
assured by the linearisation leading to~(\ref{eq:det}): the zero
eigenvalue assures that $\eta\approx\hh$ evolves nonlinearly; the $-1$
eigenvalue assures us that $u\approx\uu$ decays linearly with some
nonlinear modifications; and the lack of other eigenvalues assures us
that the normal form has two algebraic constraints on the other
variables.  The condition~(\ref{eq:zgu}) ensures that $\uu=0$\,, with
$\vv=\pp=0$\,, immediately describes the exponentially attractive
centre manifold~\cite{Elphick87b}.  Consequently, the model evolution
is simply $\hh_t=\gh(\hh,\epsilon)$\,, from~(\ref{eq:nfe}), and the
projection of initial conditions is done along curves of
constant~$\hh$.

The most straightforward construction uses iteration similar to that
introduced for constructing centre manifolds~\cite{Roberts96a}.
Suppose we have some current approximation to the transformation and
the evolution; for examples, $H=V=U=P=\gh=\gu=0$ is an initial
approximation, whereas (\ref{eq:tnfh}--\ref{eq:nfu}) is an
approximation with errors~$\Ord{\epsilon^5}$.  Given some such
approximation, we seek \emph{small} corrections, denoted by primes,
that improve the current approximation: that is, for example, we seek
$u=\uu+U+U'$ for some small~$U'$ to be determined.  Substitute these
into the governing system~(\ref{seu}--\ref{sek}) and omit products of
small corrections to obtain a system of equations to solve for
corrections.  For example, the left-hand side of the
$\eta$-equation~(\ref{sek}) becomes, upon using the chain rule,
\begin{eqnarray*}
    &&\eta_t+\epsilon u\eta-v\\
    &=& \D\eta\hh\D\hh t +\D\eta\uu\D\uu t +\epsilon u\eta-v \\
    &=& \left(1+\D H\hh+\D{H'}\hh\right)(\gh+\gh')
        +\left(\D H\uu+\D{H'}\uu\right)(-\uu+\gu+\gu')
        \\&&{}
        +\epsilon(\uu+U+U')(\hh+H+H')-(\vv+V+V') \\
    &\approx&  \left(1+\D H\hh\right)\gh
        +\D H\uu(-\uu+\gu)
        +\epsilon(\uu+U)(\hh+H)-(\vv+V)
        \\&&{}
        +\gh' -\uu\D{H'}\uu -V' \\
    &=&\R{sek} +\gh' -\uu\D{H'}\uu -V'\,,
\end{eqnarray*}
where $\R{sek}$ denotes the residual of~(\ref{sek}) at the current
approximation.  Hence to choose corrections which reduce the residual
we need to solve~(\ref{eq:toykd}) below.  Similarly for the other
equations to require the solution of
\begin{eqnarray}
    \uu\D{U'}\uu-U'&=&\R{seu}+\gu'\,,\label{eq:toyud}
    \\
    \uu\D{V'}\uu-V'-P'&=&\R{sev}\,,\label{eq:toyvd}
    \\
    -V'&=&\R{sec}\,,\label{eq:toycd}
    \\
    \uu\D{H'}\uu+V'&=&\R{sek}+\gh'\,,\label{eq:toykd}
\end{eqnarray}
where $R_i$ denotes the residual of the system's
equations~(\ref{seu}--\ref{sek}) evaluated at the current
approximation.  For example, at the initial approximation
$H=V=U=P=\gh=\gu=0$\,, and using $\vv=\pp=0$\,, the residuals
\begin{equation}
    \R{seu}=\epsilon\uu^2+\epsilon^2\uu+\epsilon\hh\,,
    \quad
    \R{sev}=0\,,
    \quad
    \R{sec}=\epsilon\uu\,,
    \quad\mbox{and}\quad
    \R{sek}=\epsilon\uu\hh\,.
    \label{eq:toy1st}
\end{equation}
Now, see that the algebraic equation~(\ref{sec}) immediately gives the
$V$-correction through~(\ref{eq:toycd}), and that the dynamic
equation~(\ref{sev}) for~$v$ immediately gives the $P$-correction
through~(\ref{eq:toyvd}).  For example, in the initial iterate
$V'=-\epsilon\uu$ and $P'=0$\,.  The other two components are more subtle.
\begin{itemize}
	\item For equation~(\ref{eq:toykd}), since $V'$ is known, any term
	involving a factor~$\uu^q$ in the residual~$\R{sek}-V'$ will cause,
	via the homological operator~$\uu\D{H'}\uu$, a term with~$\uu^q/q$
	in the transformation~$H'$; for example, with $\R{sek}-V'
	=\epsilon\uu(\hh+1)$ the correction $H' =\epsilon \uu(\hh+1)$\,.
	However, any term in the residual~$\R{sek}-V'$ with no $\uu$~factor
	cannot be matched by a term in~$H'$, as $\uu\D{H'}\uu$ necessarily
	involves~$\uu$, and must instead be matched by a term in the
	evolution~$\gu'$.  Thus the transformation is found so that the
	evolution of~$\hh$ only involves~$\hh$ itself as required
    by~(\ref{eq:nfe}).

	\item For equation~(\ref{eq:toyud}) any term involving a
	factor~$\uu^q$ in the residual~$\R{seu}$ will cause, via the
	homological operator~$\uu\D{U'}\uu-U'$, a term with~$\uu^q/(q-1)$
	in the transformation~$U'$; for example, with
	$\R{seu}=\epsilon\uu^2+\epsilon\hh$ the correction
	$U'=\epsilon\uu^2-\epsilon\hh$\,.  However, any term linear
	in~$\uu$, such as the term~$\epsilon^2\uu$ in~(\ref{eq:toy1st}),
	must be used to correct the evolution via~$\gu'$.  Thus the
    transformation is done so that the evolution of~$\uu$
    satisfies~(\ref{eq:zgu}).
\end{itemize}

In this manner iteration builds asymptotic approximations to the normal
form transformation and the evolution, such as the
earlier~(\ref{eq:tnfh}--\ref{eq:nfy}).

\subsection{Homogeneous solutions provide freedom}
\label{Shomo}

There are extra complications because the normal form transformation is
not unique.  In general the transformations~(\ref{eq:nft}) are chosen
to minimise the number of terms in the corresponding normal form
dynamical equations~(\ref{eq:nfe}) by placing as many terms as possible
in the transformation~(\ref{eq:nft}).  However, the result then depends
upon our choice of basis for the algebra.  Here we briefly explore the
options available to us while retaining all the properties required to
extract the centre manifold model and the projection of initial
conditions.

As an aside note that the essential properties of the normal form are
displayed in Figure~\ref{fig:seiso}: the curve of $\uu=0$ must
correspond to the centre manifold (the thick curve); and the curves of
constant~$\hh$ must correspond to the isochrons (thin contours).
Freedom comes from being able to parametrise these curves in state
space in any smooth way consistent with these requirements.

The simplest way to identify freedom is to find the homogeneous
solutions of the correction equations~(\ref{eq:toyud}--\ref{eq:toykd})
for~$H$, $V$, $U$, $P$, $\gh$~and~$\gu$.  The full homological
operator~\cite[e.g.]{Murdock03} for this problem appears on the
left-hand side of~(\ref{eq:toyud}--\ref{eq:toykd}); it operates on the
space of multinomials in~$\uu$, $\hh$~and~$\epsilon$, but only the
polynomials in~$\uu$ are significant as $\hh$~and~$\epsilon$ do not
appear in the operator.  First see that homogeneous solutions must have
$V'=P'=0$\,.  Then, second, two families of linearly independent
solutions are
\begin{eqnarray}&&
    (H',\gh')=(\uu^q,q\uu^q)\,,\quad U'=\gu'=0\,,
    \label{eq:toyhh2}
    \\\mbox{and}&&
    (U',\gu')=(\uu^q,(q-1)\uu^q)\,,\quad H'=\gh'=0\,.
    \label{eq:toyhh1}
\end{eqnarray}
Thus we can always change an iterate by some linear combination of
these solutions, multiplied by some function of $\hh$~and~$\epsilon$,
and only affect higher-order terms in the transformation.
\begin{itemize}
    \item  Consider~(\ref{eq:toyhh2}) with $q\geq 1$\,: introducing
    any such component will destroy the essential property that the
    evolution in~$\hh$ is independent of~$\uu$ that we require to use 
    the normal form to project initial conditions.  Thus we cannot
    use this freedom.

	\item Consider~(\ref{eq:toyhh2}) with $q=0$\,, that
	is~$(H',\gh')=(1,0)$\,: introducing such a component, multiplied by
	an arbitrary function of $\hh$~and~$\epsilon$, allows us to widely
	vary the relationship between the original variable~$\eta$ and the
	transformed variable~$\hh$.  If we maintain the exact identity
	between these variables, $\eta=\hh$ when $\uu=0$\,, then the model
	variable~$\hh$ has the same physical meaning as the
	variable~$\eta$.  Although not essential, this seems
	desirable as it enhances physical interpretation.  We choose this
	option.
    
	However, the normal form transformation may instead be used to
	immediately put the evolution equations on the centre manifold in
	one of the well established canonical normal
	forms~\cite[Chapt.~6]{Murdock03}, \cite[Chapt.~3]{Rand87}
	or~\cite[Chapt.~3]{Kuznetsov95}.  This alternative also uses the
	freedom identified here, but reduces the physical connection and 
    is not implemented here.
    
	\item Consider~(\ref{eq:toyhh1}) with $q=0$\,, that
	is~$(U',\gu')=(1,-1)$\,: we cannot use this freedom as, by
	introducing terms only in $\hh$~and~$\epsilon$ into~$\gu'$, it
	destroys the requirement that $\uu=0$ is the centre manifold of the
	$\uu$~evolution.
    
	\item Consider~(\ref{eq:toyhh1}) with $q\geq1$\,: these homogeneous
	solutions allows us to alter the precise meaning of the transformed
	variable~$\uu$\,.  Note that the linearised dynamics about the
	centre manifold $\uu=0$ are unaffected by any such change as terms
	linear in~$\uu$, corresponding to $q=1$\,, do not affect the
	evolution~$\gu'$.  For example, we could use this freedom to
	replace the strict linearity condition~(\ref{eq:zgu}) for the
	evolution of~$\uu$ and instead require the normal form
    \begin{equation}
        u=\uu+U(\hh,\epsilon)\,,\quad\mbox{and}\quad
        \gu(\hh,0,\epsilon)=0\,,
        \label{eq:nfreq}
    \end{equation}
	so that $\uu$ measures precisely the departure of the state of the
	system from the centre manifold.  For example, here the
	corresponding alternative normal form transform is
\begin{eqnarray}
    \eta&=&\hh +\epsilon(1+\hh)\uu +\epsilon^2\half (1+\hh)\uu^2 
     \nonumber\\&&\phantom{\hh}
    -\epsilon^3\rat16(1+\hh)(12-\uu^2+6\hh)\uu
    +\Ord{\epsilon^4}\,,
    \label{eq:atnfh}
    \\
    u&=&\uu- \epsilon \hh +\epsilon^3(3+\hh)\hh 
    +\Ord{\epsilon^4}\,,
    \label{eq:atnfu}
    \\
    v&=&\vv-\epsilon\uu+\epsilon^2\hh 
    +\Ord{\epsilon^4}\,,
    \label{eq:atnfv}
    \\
    p&=&\pp-\epsilon^2\hh -\epsilon^3(1+\hh)\uu
    +\Ord{\epsilon^4}\,,
    \label{eq:atnfp}
\end{eqnarray}
with corresponding evolution
\begin{eqnarray}&&
    \hh_t=\epsilon^2(1+\hh)\hh
     +\Ord{\epsilon^4}\,,
     \label{eq:anfh}
     \\&&
    \uu_t=-[1+\epsilon^2(2+\hh)]\uu -\epsilon^3\half(1+\hh)\uu^2
    +\Ord{\epsilon^4}\,,
    \label{eq:anfu}
    \\&&
    \vv=\pp=0\,.
    \label{eq:anfy}
\end{eqnarray}
We adopt this alternative when constructing the normal form of the
dynamics of thin fluid films.

\end{itemize}
Using such freedom we find the centre manifold and the projection of
initial conditions is the same.  For example, the direction of the
projection of initial conditions, the tangent to the
isochrons~\cite{Guckenheimer75, Cox93b}, at the centre manifold $\uu=0$
is the same, namely
\begin{equation}
    \bmatrix \eta_\uu \\ u_\uu \\ v_\uu \\ p_\uu\endbmatrix
    =\bmatrix \epsilon(1+\hh)-\epsilon^3(1+\hh)(2+\hh)
    \\ 1
    \\ -\epsilon
    \\ -\epsilon^3(1+\hh) \endbmatrix +\Ord{\epsilon^5}\,.
\end{equation}
This equivalence is reassuring because the different possible
transformations still represent the same dynamical processes.

\section{Normal form of thin fluid film equations}
\label{Sflow}

In this section we take the equations for a thin fluid film and
transform them into a normal form that separates the dynamics of the
viscously decaying modes from the large scale mode of slow evolution 
of the fluid film's thickness.

\subsection{Nondimensionalise the fluid film equations}

Consider the two-dimensional flow of a thin film of Newtonian fluid
along a flat substrate.  We adopt a nondimensionalisation based upon
the characteristic thickness of the film~$H$, and some characteristic
velocity~$U$: for a specific example, in a regime where surface tension
drives a flow against viscous drag the characteristic velocity
is~$U=\sigma/\mu$ and thus the Weber number~$\we=\sigma/(U\mu)=1$\,;
alternatively~$U$ might be chosen to characterise the difference between
a given initial lateral velocity and that corresponding to the
lubrication model~(\ref{eq:cmmod}).  Reverting to the general case, the
reference length is~$H$, the reference time~$H/U$, and the reference
pressure~$\mu U/H$.  The varying free surface is located at
$y=\eta(x,t)$, where $x$~and~$y$ are lateral and normal coordinates
respectively.  The flow, with velocity~$\vel=(u,v)$ and pressure~$p$,
is governed by the nondimensional incompressible Navier-Stokes
equations
\begin{eqnarray}
        \re(\vel_t+\vel\cdot\grad\vel)
        & = & -\grad p+\grad^2\vel\,,\label{nseq}
\end{eqnarray}
where~$\re=UH\rho/\mu$ is a Reynolds number characterising the
importance of the inertial terms compared to viscous dissipation,
supplemented by the continuity equation
\begin{eqnarray}
        \divv\vel & = & 0\,,
\label{cont}
\end{eqnarray}
non-slip boundary conditions on the substrate
\begin{equation}
        \vel=\vec 0\quad\mbox{on $y=0$\,,}
        \label{svebbc}
\end{equation}
and tangential stress and normal stress conditions on the free surface
\begin{eqnarray}
  (1-\eta_x^2)(u_y+v_x)=2\eta_{x}(u_x-v_y)
  \quad\mbox{on $y=\eta$\,,}
  \label{svetc}&&\\
    p+\frac{\we\eta_{xx}}{(1+\eta_x^2)^{3/2}}
    =\frac2{1+\eta_x^2}\left[ v_y+u_x-\eta_x(u_y+v_x)
    \right]
    \quad\mbox{on $y=\eta$\,.}&&
        \label{svenc}
\end{eqnarray}
We close the problem with the kinematic condition relating the 
velocity of the fluid on the surface to the evolution of the free 
surface:
\begin{equation}
     \eta_t = -\pdx \int_0^\eta u\,dy
     =v-u\eta_x\quad\mbox{on $y=\eta$\,.}
        \label{svekc}
\end{equation}
The fluid film is assumed to be so thin that the gravitational force in
the momentum equations is neglected in this initial research project.
The other main assumption we make is that the lateral variations are 
slow.

\subsection{Introduce the normal form transform}

Centre manifold techniques construct a model of the fluid flow assuming
that lateral derivatives, $\pdx$, are small~\cite[e.g.]{Roberts96a,
Roberts03b}.  Here we limit our aim to encompass flows with a lateral
velocity which is also a small departure from the centre manifold of
slow evolution.  In the normal form we thus seek fields $\uu(x,y,t)$,
$\vv(x,y,t)$, $\pp(x,y,t)$ and~$\hh(x,t)$ which are a near identity
transform of the physical fluid fields.  We label these ``quasi-''
because they are approximately the same as the well known physical
fields.  Based upon the flow states we know for thin fluid films:
\begin{eqnarray}&&
    \eta= \hh+\Oiv\,,
    \label{eq:thnf}\\&&
    v = \vv+\Oiv\,,
    \label{eq:tvnf}\\&&
    p = \pp-\we\hh_{xx}+\Oiv\,,
    \label{eq:tpnf}\\&&
    u = \uu+\we(\hh y-\half y^2)\hh_{xxx}+\Ord{\pdx^5}\,.
    \label{eq:tfnf}
\end{eqnarray}
As in Section~\ref{Shomo}, see equation~(\ref{eq:nfreq}), we have here
defined the lateral quasi-velocity~$\uu$ to be exactly the lateral
velocity field's departure from that of the centre manifold model,
approximately $\we(\hh y-\half y^2)\hh_{xxx}$\,.  Because of the
algebraic components in the fluid film equations we know the normal
form has the two trivial algebraic equations
\begin{equation}
    \vv=0\quad\mbox{and}\quad\pp=0\,.
    \label{eq:trnf}
\end{equation}
Initially we assume the normal form also has trivial algebraic boundary
conditions:
\begin{equation}
    \uu=\vv=0\mbox{ on }y=0\,,\quad
    \D \uu y=\pp=0\mbox{ on }y=\hh\,;
    \label{eq:fsnf}
\end{equation}
but later we find it is necessary to modify one of these.  Lastly, the
lateral quasi-velocity and the fluid quasi-thickness evolve according
to
\begin{eqnarray}
    \D\uu t&=&\gu(\uu,\hh)=\frac1\re\DD\uu y +\Ov\,,
    \label{eq:gunf}\\
    \D\hh t&=&\gh(\hh)=-\third(\hh^3\hh_{xxx})_x+\Ord{\pdx^6} \,,
    \label{eq:ghnf}
\end{eqnarray}
where $\gu(0,\hh)=0$\,.  This last property of~$\gu$ ensures that the
trivial $\uu=0$ forms the centre manifold of the dynamics.
Hence~(\ref{eq:gunf}) encapsulates that lateral shear flows dominantly
decay by viscous diffusion, and the classic thin film model,
$\hh_t\approx -\third(\hh^3\hh_{xxx})_x$\,, is then recovered.  This is
our purpose for the normal form
transformation~(\ref{eq:thnf}--\ref{eq:tfnf}).

Note that the very small difference, $\Oiv$, between the physical fluid
thickness~$\eta$ and the quasi-thickness~$\hh$ identified
in~(\ref{eq:thnf}) means that evaluation on $y=\eta$ and $y=\hh$ are
almost everywhere interchangeable to the order of accuracy determined
in this section.
  
Also note that the form of many of the order of errors in the above are
equivalent to having just one scaling parameter and assuming the
lateral quasi-velocity field~$|\uu|$ scales
with~$\partial_x^3$.\footnote{I use the notation that an asymptotic
error~$\Ord{\alpha^p+\beta^q}$ to denote the error may involve terms
in~$\alpha^m\beta^n$ for $p/m+q/n\geq 1$\,.  Thus, for example, here
with small quantities $\uu$~and~$\pdx$, $\uu_{xx}$, $\hh_x^2\uu$
and~$\hh_x^2\hh_{xxx}$ are all~$\Ov$.} However, here we are primarily
interested in determining the dominant nontrivial effects of a lateral
velocity field and thus we do not seek higher order in~$\pdx$ except
where $\uu$~is involved.  Consequently we later move to refine the
approximations to a form where~$|\uu|$ is asymptotically larger and
scales with~$\pdx^2$---adopting a single scaling parameter would hinder
this development so we adopt this more flexible expression of
asymptotic errors.

In principle we proceed to use an iteration process to deduce the
details of the normal form transformation.  However, for the moment we
only attempt just a little more than the first iteration.  This is
enough to discover highly nontrivial properties of the projection of
initial conditions.

\subsection{Fluid conservation evolves the free surface} 
\label{SScon}

Seek a correction to the free surface $\eta=\hh+\eta'$\,, using
conservation of fluid~(\ref{svekc}):\footnote{The same result is also
obtained from the kinematic condition, $\eta_t=v-u\eta_x$ on
$y=\eta$\,, using the next approximation for the normal velocity field
which is obtained in the next subsection.}
\begin{eqnarray}&&
    \D\eta t= -\pdx \int_0^\eta u\,dy
    \nonumber\\&\Rightarrow& 
    \D\hh t+\D{\eta'}t=
    -\pdx\left[\int_0^\eta \uu\,dy
    +(\half\hh\eta^2-\rat16\eta^3)\hh_{xxx} \right]
    +\Ovi
    \nonumber\\&\Rightarrow& 
    \D\hh t+\D{\eta'}t=
    -\pdx\int_0^\hh \uu\,dy -\third(\hh\hh_xxx)_x
    +\Ovi
    \nonumber\\&\Rightarrow& 
    \D{\eta'}t=-\pdx\int_0^\hh \uu\,dy
    +\Ovi\,.
    \label{eq:etadnf}
\end{eqnarray}
We used the known evolution of the film thickness~(\ref{eq:ghnf}), that
$\hh_t\approx -\third(\hh^3\hh_{xxx})_x$\,; if we had not known this
already, then at this step we would have discovered it was necessary;
we can only put into~$\eta'$ terms which involve~$\uu$ because terms
which only involve~$\hh$ are rendered ineffective in~$\D{\eta'}t$ by
the very slow evolution of~$\hh$.  Due to the form of the
right-hand side for $\D{\eta'}t$ in~(\ref{eq:etadnf}), now try a change
to the fluid thickness involving a weighted integral of the lateral
velocity:
\begin{eqnarray*}&&
    \eta'= \pdx \int_0^\hh  w(y;\hh)\uu \,dy
    +\Ovi
    \\&\Rightarrow& 
    \D{\eta'}t=
    \pdx \Bigg\{ \underbrace{\D\hh t\uu(\hh)w(\hh;\hh)}_{\text{negligible}} 
    +\int_0^\hh w\D\uu t 
    +\underbrace{\D w\hh\D\hh t\uu}_{\text{negligible}}\,dy \Bigg\}
    +\Ovi
    \\&\Rightarrow& 
    \D{\eta'}t=
    \pdx\left\{ \int_0^\hh w\frac1\re\DD\uu y\,dy \right\}
    +\Ovi
    \\&\Rightarrow& 
    \D{\eta'}t=
    \pdx\left\{ \frac1\re\left[ w\D\uu y-w'\uu \right]_0^\hh+
    \int_0^\hh w''(y)\frac1\re\uu\,dy \right\}
    +\Ovi\,,
\end{eqnarray*}
where primes on the weight function~$w$ denote $y$~derivatives.
Equate this to~(\ref{eq:etadnf}), recognising that $\uu=0$ on $y=0$
and $\D\uu y=0$ on $y=\hh$\,, to deduce the weight function satisfies
\begin{equation}
    w''=-\re\,,\quad w(0)=0\,,\quad w'(\hh)=0
    \quad\Rightarrow\quad
    w=\re(\hh y-\half y^2)\,.
\end{equation}
Thus the free surface 
\begin{equation}
    \eta=\hh+\re\pdx\int_0^\hh (\hh y-\half y^2)\uu \,dy 
    +\Ovi\,,
    \label{eq:etanf}
\end{equation}
in terms of the normal form fields $\hh$~and~$\uu$.

\paragraph{Initial conditions:}
Revert~(\ref{eq:etanf}) and~(\ref{eq:tfnf}) to
\begin{displaymath}
    \hh\approx \eta-\re\pdx\int_0^\eta (\eta y-\half y^2)u \,dy
    \,,\quad\mbox{and}\quad
    \uu\approx u-(\eta y-\half y^2)\eta_{xxx}\,.
\end{displaymath}
Then recall that in the normal form~(\ref{eq:gunf}--\ref{eq:ghnf}), the
quasi-velocity $\uu\to0$ quickly through viscosity.  Thereafter, when
$\uu=0$\,, the fluid thickness~$\eta$ and the quasi-thickness~$\hh$ are
identical, by~(\ref{eq:etanf}).  Further, $\eta$~and~$\hh$ evolve
according to the same models, (\ref{eq:cmmod})~and~(\ref{eq:ghnf})
respectively, and in the normal form~(\ref{eq:gunf}--\ref{eq:ghnf}) the
evolution of the quasi-thickness~$\hh$ is independent of
quasi-velocity.  Hence, from the above reversion, to make the correct
long term forecast with the lubrication model we must start with
the initial fluid thickness~$\eta(x,0)$ as claimed in~(\ref{eq:cmic}).

In the remainder of this section we demonstrate that this initial
condition is the first nontrivial approximation in an asymptotic
expansion based upon the normal form of thin film fluid dynamics.

\subsection{Continuity updates the normal velocity} 
Upon substituting the current approximation, $u$ from~(\ref{eq:tfnf})
and $v\approx\vv+v'$ where $\vv=0$ from~(\ref{eq:trnf}), the continuity
equation~(\ref{cont}) implies
\begin{equation}
    v = \vv-\int_0^y \uu_x\,dy-\pdx [
    (\half\hh y^2 -\rat16 y^3)\hh_{xxx}] +\Ovi\,.
\end{equation}
As necessary, for quasi-fields $\vv=\uu=0$, this reduces to the well
established normal velocity field for the centre manifold model.

\subsection{Normal momentum updates pressure} 

Seek to update $p\approx\pp-\we\hh_{xx}+p'$.  Noting $u\D vx,v\D
vy=\Oviii$ and $\DD vx=\Ovi$\,, consider
\begin{eqnarray*}
    &&\re\left[\D vt+u\D vx+v\D vy\right]+\D py =\DD vx+\DD vy
    \\&\Rightarrow&
    -\uu_{xy} +\uu_{xy}^0 +\D{p'}y = -\uu_{xy} -\we\pdx[(\hh-y)\hh_{xxx}]
    +\Ovi
    \\&\Rightarrow& p'= \we\pdx[\half(\hh-y)^2\hh_{xxx}]
    +(\hh-y)\uu_{xy}^0 +p'' +\Ovi\,,
\end{eqnarray*}
where the superscript~$0$ on~$\uu_{xy}^0$ denotes evaluation on the
substrate~$y=0$\,, soon the superscript~$\hh$ will denote evaluation
on the fluid surface~$y=\hh$\,, and $p''(x,t)$ is an integration
constant to be determined from the free surface normal stress.  The
normal stress condition on $y=\eta$ is equivalent to evaluating on
$y=\hh$ to as yet negligible relative error of~$\Oiv$\,; thus
\begin{eqnarray*}
    &&p(1+\eta_x^2)+\we\eta_{xx}(1-\half \eta_x^2) 
    \\&&{}
    =2\left[ \D vy +\eta_x^2\D ux -\eta_x\left(\D uy
    +\D vx \right) \right] +\Ovi
    \\&\Rightarrow&
    -\we(\hh_{xx}+\hh_x^2\hh_{xx})+p''
    +\we(\hh_{xx}-\half \hh_x^2\hh_{xx})
    \\&&{}
    = 2\left[ -\uu_x^\hh-\half\pdx(\hh^2\hh_{xxx})
    +0 -\hh_x\uu_y^\hh -0 \right] +\Ovi
    \\&\Rightarrow&p'' = \we\pdx(\hh^2\hh_{xxx}) +\we\rat32\hh_x^2\hh_{xx}
    -2\uu_x^\hh -2\uu_y^\hh\hh_x +\Ovi\,,
\end{eqnarray*}
Hence the more refined description of the pressure field is
\begin{eqnarray}
    p &=& \pp -\we\left\{\hh_{xx} -\rat32\hh_x^2\hh_{xx}
    +\half\pdx[(\hh^2+2\hh y-y^2)\hh_{xxx}]\right\}
    \nonumber\\&&{}
    +(\hh-y)\uu_{xy}^0 -2\pdx(\uu^\hh) +\Ovi\,.
\end{eqnarray}

\subsection{Lateral momentum determines lateral velocity}

Seek to update $u\approx\uu+u'$ and $\D\uu t\approx\frac1\re\DD\uu
y+\gu'$\,.  However, as flagged earlier, we are more interested in the
effects of the lateral velocity field~$\uu$ in the normal form rather
than higher orders in the lateral gradients~$\pdx$ modifying the
leading order evolution~(\ref{eq:ghnf}).  Thus we change our
expressions of errors to a form equivalent to scaling the lateral
velocity~$\uu$ with~$\pdx^2$ rather than with~$\pdx^3$ as done so far.
Then working to errors~$\ov$ enables us to resolve terms such as
lateral diffusion~$\uu_{xx}$ while omitting fifth derivatives of the
free surface.

Using computer algebra, see~\cite{Roberts04a} for the \textsc{Reduce}
code, which also confirms that the residual in the Navier-Stokes
equations~(\ref{nseq}) is~$\ov$, we determine
\begin{eqnarray}
    &&\re\left[\D ut +u\D ux +v\D uy \right]+\D px =\DD ux+\DD uy
    \nonumber\\&\Rightarrow& \re\gu'+\re\D{u'}t-\DD{u'}y
    =\uu_{xx}
    -\pdx[(\hh-y)\uu_{xy}^0] 
    \nonumber\\&&\quad{}
    +2\uu_{xx}^\hh -2\hh_x^2\uu_{yy}^\hh
    +\ov\,,
    \label{eq:uc5}
\end{eqnarray}
At the free surface the tangential stress supplies
\begin{eqnarray}
    &&(1-\eta_x^2)\left(\D uy+\D vx\right)
    +2\eta_x\left( \D vy+\D ux \right)=0
    \nonumber\\&\Rightarrow& \D{u'}y =\int_0^\hh \uu_{xx}\,dy
    +4\uu_x^\hh\hh_x
    +\ov\,,
    \label{eq:ufsbc}
\end{eqnarray}
when evaluated on the free surface, $y\approx\hh$\,.  We could satisfy
this stress condition by changing the velocity field in the interior
through a component in~$u'$.  However, recall that the lateral
quasi-velocity~$\uu$ has more direct meaning when the normal form is
chosen so that the physical velocity~$u$ is independent of~$\uu$ except
for the leading term, see~(\ref{eq:tfnf}).  Thus, satisfy this
inhomogeneous \textsc{bc} by changing the surface boundary
condition~(\ref{eq:fsnf}) for~$\uu$ to
\begin{equation}
    \D\uu y=\int_0^\hh \uu_{xx}\,dy +4\hh_x\uu_x^\hh +\ov 
    \quad\mbox{on }y=\hh\,.
    \label{eq:uubc1}
\end{equation}
Consequently a new term on the left-hand side of~(\ref{eq:ufsbc})
cancels the right-hand side terms, leaving the homogeneous \textsc{bc}
$\D{u'}y=0$ on $y=\hh$\,.  Now consider the lateral momentum update
equation~(\ref{eq:uc5}): all terms in the right-hand side involve~$\uu$
and so they are all placed into the evolution~$\gu'$.  Thus the lateral
velocity is still given by~(\ref{eq:tfnf}), but we improve the
description of the evolution~to
\begin{equation}
     \re\D \uu t =\uu_{yy}+\uu_{xx}
    -\pdx[(\hh-y)\uu_{xy}^0]
    +2\uu_{xx}^\hh -2\hh_x^2\uu_{yy}^\hh
    +\ov\,.
     \label{eq:uut1}
\end{equation}
As required, the normal form equation~(\ref{eq:uut1}) with boundary
condition~(\ref{eq:uubc1}) has $\uu=0$ as an attractive invariant
(centre) manifold.  Then with lateral quasi-velocity $\uu=0$ the
lateral velocity field~(\ref{eq:tfnf}) reduces to the conventional
thin fluid film approximation $u\approx\we(\hh y-\half y^2)\hh_{xxx}$\,.

\subsection{Fluid conservation refines the evolution}

We seek a further refinement to the description of the fluid
thickness~(\ref{eq:etanf}) using conservation of fluid~(\ref{svekc}).
The aim is to discover more effects of the lateral velocity and so we
work to errors~$\ovi$.  There is considerable detail in determining the
new terms, relegated to the computer algebra in~\cite{Roberts04a},
but the basic technique follows that used in~Section~\ref{SScon}.  To the
required order of accuracy we know the quasi-thickness
evolution~(\ref{eq:ghnf}); thus the only freedom available is to update
the fluid thickness from~(\ref{eq:etanf}) by some small change~$\eta''$.
Substitute the lateral velocity field into~(\ref{svekc}) to obtain
(using superscripts $0$~and~$\hh$ to denote evaluation on the bed and
quasi-surface, but now also using superscript~$y$ to denote
$\int_0^y\cdot\,dy$)
\begin{eqnarray}
    \D{\eta''}t&=&
    +\rat18\hh^4\uu_{xxxy}^0
    +\rat56\hh^3\hh_x\uu_{xxy}^0
    +\rat13\hh^3\hh_{xx}\uu_{xy}^0
    +\hh^2\hh_x^2\uu_{xy}^0
    -\rat23\hh^3\uu_{xxx}^\hh
    \nonumber\\&&{}
    +2\hh^3\hh_x^2\uu_{xyy}^\hh
    +2\hh^3\hh_x\hh_{xx}\uu_{yy}^\hh
    -\hh^2(\uu_{xxx}^y)^\hh
    -5\hh^2\hh_x\uu_{xx}^\hh
    -2\hh^2\hh_{xx}\uu_x^\hh
    \nonumber\\&&{}
    +4\hh^2\hh_x^3\uu_{yy}^\hh
    -2\hh\hh_x(\uu_{xx}^y)^\hh
    -4\hh\hh_x^3\uu_x^\hh
    +(\uu_{xxx}^{yyy})^\hh
    +\hh_x(\uu_{xx}^{yy})^\hh
    \nonumber\\&&{}
    +\ovi\,.
    \label{eq:etadnfo}
\end{eqnarray}
All these residual terms are~$\Ord{u\pdx^3}$, there is no component
of~$\Ord{u\pdx^2}$.  We summarise the details of the derivation in the
next paragraph, but a solution for the above refinement of the fluid 
thickness~$\eta$ is
\begin{eqnarray}
    \eta''(\hh,\uu)&=&\re\Big\{
\tint_0^\hh (\rat{1}{24}y^4 -\rat{1}{6}\hh y^3 -\rat{1}{4}\hh^2y^2 
+\rat{3}{2}\hh^3y -\rat{1}{8}\hh^4)\uu_{xxx}\,dy
    \nonumber\\&&{}
+\hh_{x}\tint_0^\hh (-\rat{1}{6}y^3-\rat{1}{2}\hh y^2+\rat{13}{2}\hh^2y
-\rat{5}{6}\hh^3)\uu_{xx}\,dy
    \nonumber\\&&{}
+\hh\hh_{x}^2\tint_0^\hh (4y-\hh)\uu_{x}\,dy
+\hh^2\hh_{xx}\tint_0^\hh (2y-\rat{1}{3}\hh)\uu_{x}\,dy
    \nonumber\\&&{}
+12\hh^3\hh_{x}^2 \uu_{x}^\hh
+\rat{26}{3}\hh^3\hh_{x}\hh_{xx} \uu^\hh
+7\hh^2\hh_{x}^3 \uu^\hh
+3\hh^4\hh_{x} \uu_{xx}^\hh
    \nonumber\\&&{}
+2\hh^4\hh_{xx} \uu_{x}^\hh
+\hh^4\hh_{xxx} \uu^\hh
-2\hh^4\hh_{x}^3 \uu_{yy}^\hh   
    \Big\}\,.
    \label{eq:refineh}
\end{eqnarray}
Thus for this~$\eta''$ the normal form for the fluid thickness is now
\begin{equation}
    \eta=\hh+\re\pdx\tint_0^\hh (\hh y-\half y^2)\uu \,dy 
    +\eta''(\hh,\uu)+\ovi\,.
    \label{eq:etanfo}
\end{equation}
In the next section we use this transformation to determine more
details about the initial conditions for the model of thin fluid
film~(\ref{eq:cmmod}).

But before proceeding we overview the machinations needed to derive the
refinement~(\ref{eq:refineh}).  Recall that the left-hand side $\D
{\eta'}t$ is approximated by a formal expression:
\begin{displaymath}
    \D{\eta'}t\approx \frac1\re\D{\eta'}\uu\DD{\uu}y\,.
\end{displaymath}
Thus the components in the residual on the right-hand side
of~(\ref{eq:etadnfo}) which involve two or more derivatives of~$y$ can
be matched by a component in the refinement~$\eta''$ with two less
derivatives in~$y$.  The other components, involving the first
$y$-derivative or integrals of the quasi-velocity~$\uu$, must be
matched by an integral component in the refinement~$\eta''$ as we did
in~Section~\ref{SScon}.  I achieve such matching through considerable trial
and error, and in three stages.
\begin{itemize}
	\item Starting with the term with the most $y$-integrals,
	namely~$(\uu_{xxx}^{yyy})^\hh$, the three $y$-derivatives in this
	boundary contribution is matched through multiple integration by
	parts of $\partial_t\uu_{xx}\propto\uu_{xxyy}$ with a weight
	function which is quartic in~$y$.  This and the bed component
	in~$\hh^4\uu_{xxxy}^0$ gives the first integral
	in~(\ref{eq:refineh}).  This first term in~(\ref{eq:refineh}) also
	changes some other terms in the residual~(\ref{eq:etadnfo}).  Then
	we progress to the other terms with less $y$-integrals and match
	them with simpler integrals as seen in~(\ref{eq:refineh}).

	\item When all components as yet unmatched have two or more
	$y$-derivatives then we simply directly match them with the
	boundary contributions seen in~(\ref{eq:refineh}).

	\item However, there is a difficulty: terms with odd
	$y$~derivatives and evaluated at the quasi-surface cannot be
	matched in the above procedure.  Instead we use the surface
	boundary condition~(\ref{eq:uubc1}), either directly or its
	derivatives, to eliminate such terms.  For example,
	differentiating~(\ref{eq:uubc1}) with respect to~$x$ gives
    \begin{eqnarray*}
        &&\uu_{xy}^\hh+\hh_x\uu_{yy}^\hh
        \\
        &=&\tint_0^\hh\uu_{xxx}\,dy
        +5\hh_x\uu_{xx}^\hh
        +4\hh_{xx}\uu_x^\hh
        +4\hh_x^2\uu_{xy}^\hh+\ov
        \\
        &=&\tint_0^\hh\uu_{xxx}\,dy
        +5\hh_x\uu_{xx}^\hh
        +4\hh_{xx}\uu_x^\hh
        -4\hh_x^3\uu_{yy}^\hh
        +\ov\,,
    \end{eqnarray*}
	where the last term~$\uu_{xy}^\hh$ has a single $y$-derivative and
	so is replaced by its leading order approximation, namely
	$-\hh_x\uu_{yy}^\hh$.  The above and higher order derivatives are
	implemented in computer algebra~\cite{Roberts04a}.  Similarly, time
	derivatives of~(\ref{eq:uubc1}), recalling $\re\uu_t=\uu_{yy}$\,,
	provide replacements for higher order odd $y$-derivatives of~$\uu$.
\end{itemize}
With these methods we find that~(\ref{eq:refineh}) is the solution
for the update~$\eta''$.

\section{Revert the normal form to give initial conditions}
\label{Srev}

Given an initial state of the fluid film we determine the appropriate
initial condition for the fluid thickness in the lubrication
model~(\ref{eq:cmmod}).  First translate the initial condition to the
normal form variables $\hh$~and~$\uu$, then since the evolution
of~$\hh$ is independent of quasi-velocity~$\uu$, the correct initial
condition for the lubrication model~(\ref{eq:cmmod}) is
$\eta(x,0)=\hh_0$\,.

Suppose the fluid has initial thickness~$\eta_0(x)$, lateral velocity
field~$u_0(x,y)$, normal velocity field~$v_0(x,y)$ and pressure
field~$p_0(x,y)$.  The fields $v_0$~and~$p_0$ must be consistent with
the fluid equations~(\ref{nseq}--\ref{svekc}), but then play no role in
the following.  We revert~(\ref{eq:tfnf}) to obtain the initial
quasi-lateral velocity
\begin{equation}
    \uu_0=u_0-\we(\eta_0y-\half y^2)\eta_{0xxx}+\ovi\,,
    \label{eq:uuz}
\end{equation}
as the quasi-thickness~$\hh=\eta+\oiii$, by~(\ref{eq:etanfo}) for
example, and so $\hh$~is replaced by the initial fluid
thickness~$\eta_0$ to our order of accuracy.  This quasi-lateral
velocity~$\uu_0$ characterises the distance from the initial fluid
state to a state of the lubrication model.

Now find how this distance affects the initial fluid thickness.
Revert~(\ref{eq:etanfo}) and evaluate at the initial fluid state,
using~(\ref{eq:uuz}) and $\hh=\eta+\oiii$\,, to give the initial
quasi-thickness
\begin{equation}
    \hh_0=\eta_0-\re\pdx\int_0^{\eta_0} (\eta_0 y-\half y^2)\uu_0 \,dy 
    -\eta''(\eta_0,\uu_0)+\ovi\,,
    \label{eq:etanfoz}
\end{equation}
where $\eta''$ is specified by~(\ref{eq:refineh}).  Recall the purpose
of the normal form is to ensure the evolution of quasi-thickness~$\hh$
is independent of the quasi-velocity and hence the transient viscous
decay of lateral velocity will bring the fluid to \emph{the} solution
of the lubrication model~(\ref{eq:cmmod}) which started from the
initial condition $\eta(x,0)=\hh_0$ as specified in~(\ref{eq:etanfoz}).
This initial condition permits us to make accurate long term forecasts
with the model.

\section{Conclusion}

We use a normal form transformation to illuminate the dynamics of a
thin layer of fluid.  This is achieved by decoupling the slow
long-lasting lubrication mode from the viscously decaying lateral shear
modes.  A simple example shows that the differential algebraic nature
of the fluid equations is handled by straightforward modifications of
the usual procedure to construct a normal form.  Further, invoking the
slowly-varying assumption that lateral derivatives are small enables us
to deduce a normal form for the fluid dynamics albeit with significant
technical detail requiring computer algebra to check.  This normal form
then provides us with the rationale to choose the initial
condition~(\ref{eq:cmic}) to make forecasts with the lubrication
model(\ref{eq:cmmod}).  The next challenge is to connect this normal
form analysis to the projection of initial conditions, started
in~\cite{Suslov98b}, which is based upon the adjoint near a centre
manifold model.

\paragraph{Acknowledgement:} I thank Sergey Suslov for his valuable
input into all stages of this work.

\bibliographystyle{plain}
\bibliography{bib,ajr,new}

%

\end{document}